\documentclass[final,aps,prb,twocolumn,superscriptaddress,showpacs]{revtex4-1}
\usepackage{graphicx}% Include figure files
\usepackage{dcolumn}% Align table columns on decimal point
\usepackage{multirow}
\usepackage{bm,amsmath}% bold math
\usepackage{mathtools}
\usepackage{xcolor}
\usepackage{pdfcomment}
\usepackage[normalem]{ulem}

\makeatletter
\def\@fnsymbol#1{\ensuremath{\ifcase#1\or \dagger\or *\or *\or \ddagger\or \mathsection\or \mathparagraph\or \|\or **\or \dagger\dagger \or \ddagger\ddagger \else\@ctrerr\fi}}
\makeatother

\begin{document}

\title{Importance of anisotropic Coulomb interactions in the electronic and magnetic properties of Mn$_3$O$_4$}

\author{Sangmoon Yoon}
\altaffiliation[Present address: ]{Material Science and Technology Division, Oak Ridge National Laboratory, Oak Ridge, TN 37831, USA.}
\affiliation{Department of Materials Science and Engineering, Seoul National University, Seoul, 08826, Korea}
\affiliation{Department of Physics, Kyung Hee University, Seoul, 02447, Korea}

\author{Sangmin Lee}
\affiliation{Department of Materials Science and Engineering, Seoul National University, Seoul, 08826, Korea}
             
\author{Subeen Pang}
\affiliation{Department of Materials Science and Engineering, Seoul National University, Seoul, 08826, Korea}

\author{Miyoung Kim}
\email[Corresponding author. E-mail: ]{mkim@snu.ac.kr}
\affiliation{Department of Materials Science and Engineering, Seoul National University, Seoul, 08826, Korea}

\author{Young-Kyun Kwon}
\email[Corresponding author. E-mail: ]{ykkwon@khu.ac.kr}
\affiliation{Department of Physics, Kyung Hee University, Seoul, 02447, Korea}
\affiliation{Department of Information Display, Kyung Hee University, Seoul, 02447, Korea}

\date{\today}

%---------------------------------------------------------------------
\begin{abstract}
We report the importance of anisotropic Coulomb interactions in DFT$+U$ calculations of the electronic and magnetic properties of Mn$_3$O$_4$. The effects of anisotropic interactions in Mn$^{2+}$ and Mn$^{3+}$ are separately examined by defining two different sets of Hubbard parameters: $U^{2+}$ and $J^{2+}$ for Mn$^{2+}$ and $U^{3+}$ and $J^{3+}$ for Mn$^{3+}$. The anisotropic interactions in Mn$^{3+}$ have a significant impact on the physical properties of Mn$_3$O$_4$ including local magnetic moments, canted angle, spontaneous magnetic moment, and superexchange coupling, but those in Mn$^{2+}$ do not make any noticeable difference. Weak ferromagnetic interchain superexchange, observed in experiments, is predicted only if a sizable anisotropic interaction is considered in Mn$^{3+}$. By analyzing the eigenoccupations of the on-site Mn density matrix, we found that the spin channel involving Mn$^{3+}$ $d_{x^2-y^2}$ orbitals, which governs the 90$^\circ$ correlation superexchange, is directly controlled by the anisotropic interactions. These findings demostrate that the exchange correction $J$ for the intraorbital Coulomb potential is of critical importance for first-principles description of reduced Mn oxides containing Mn$^{3+}$ or Mn$^{4+}$.
\end{abstract}
%---------------------------------------------------------------------

\maketitle

\section{Introduction}
\label{Introduction}

\begin{figure}[t]
\includegraphics[width=1\linewidth]{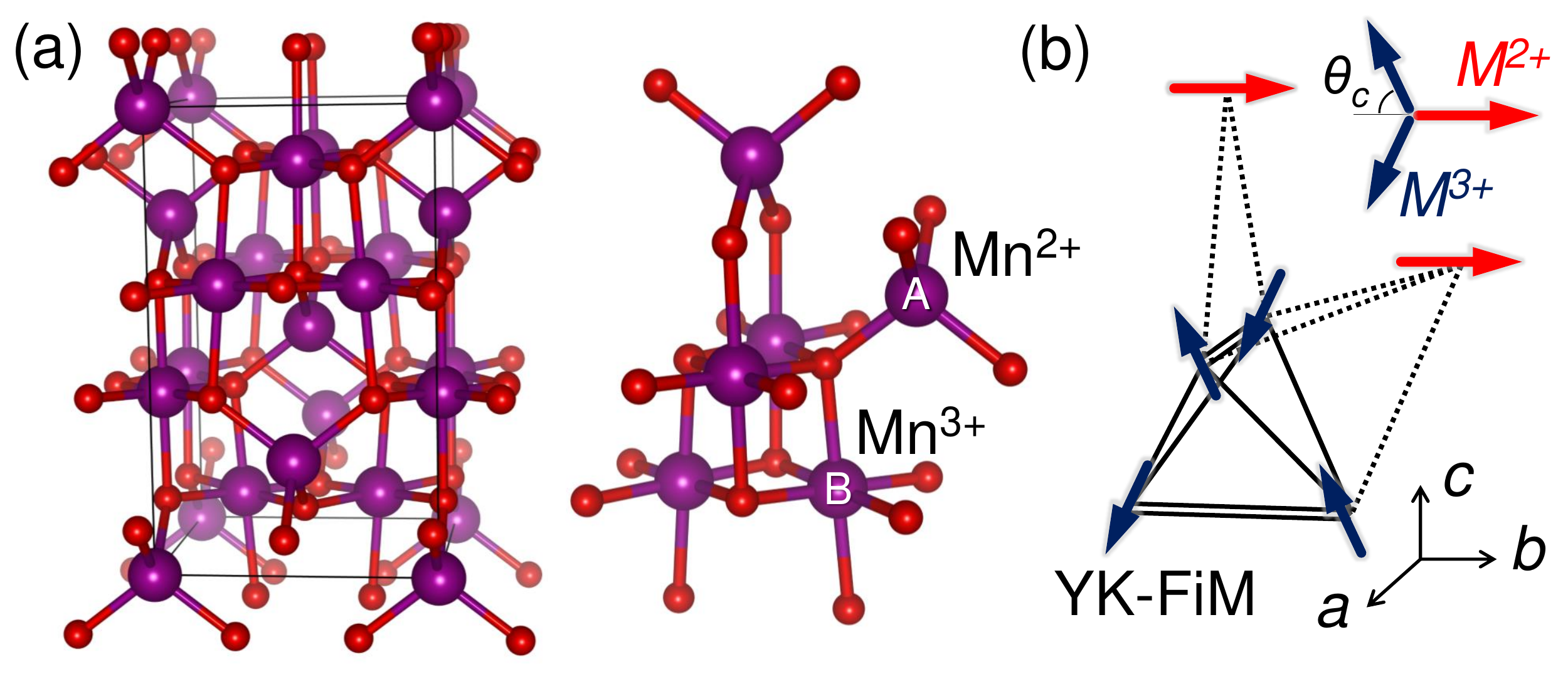}
\caption{(a) Crystal structure of the tetragonal Mn$_3$O$_4$ (left) and its local coordination with Mn$^{2+}$ occupying the tetrahedral A-site and Mn$^{3+}$ occupying the octahedral B-site (right). The purple and red balls denote Mn and O atoms, respectively. (b) Schematic of noncollinear spin configuration in the low-temperature Yafet-Kittel ferrimagnetic (YK-FiM) phase of Mn$_3$O$_4$ corresponding to the right structure in (a). The red and black arrows indicate the spins of the Mn$^{2+}$ and Mn$^{3+}$ atoms, respectively. The detailed magnetic structure of the YK-FiM phase is explained in the main text. \label{Fig1}}
\end{figure}

Mixed-valent manganese oxide Mn$_3$O$_4$ has drawn much attention because it is a prototypical example that involves Jahn-Teller distortions and geometrical frustrations as well as strong electron correlations.~\cite{{i1},{i2},{i3}} Mn$_3$O$_4$ exists in a spinel structure (AB$_2$O$_4$), with Mn$^{2+}$ and Mn$^{3+}$ respectively occupying tetrahedral sites (A sites) and octahedral sites (B sites) as shown in Fig.~\ref{Fig1}(a). It undergoes a cubic ($Fd\bar{3}m$) to tetragonal ($I4_1/amd$) structural phase transition at $T=1443$~K because of the Jahn-Teller ordering at the Mn$^{3+}$ sites.~\cite{a1} It also goes through the magnetic phase transition from paramagnetic to ferrimagnetic at its corresponding Curie temperature $T_\mathrm{C}=41$~K, adopting a noncollinear Yafet-Kittel ferrimagnetic (YK-FiM) phase with isosceles triangles, each composed of one Mn$^{2+}$ at the A site and two Mn$^{3+}$ ions at the B$_1$ and B$_2$ sites. Spins at the A sites aligns ferromagnetically along the $b$-axis, whereas spins at the B$_1$ (B$_2$) point with a canted angle of $\theta_\mathrm{c}$ toward the $+c$-axis ($-c$-axis) from the $-b$-axis as shown in Fig.~\ref{Fig1} (b).~\cite{{a2},{a3}} It has also been known that the magnetic structure of Mn$_3$O$_4$ strongly couples to its orbital, charge, and lattice, which leads to novel magnetodielectric~\cite{{i7},{i8}}, magnetoelastic~\cite{{i2},{i3},{i4},{i12}}, and magnetocaloric~\cite{i9} responses at low temperatures. The exchange couplings of Mn$_3$O$_4$ have been quantitatively estimated from the spin wave excitations measured by neutron scattering experiments, and have demonstrated that the exchange interactions are determined by the direct orbital overlap and the 90$^\circ$ superexchange coupling between two Mn$^{3+}$ ions via an oxygen bridge.~\cite{{i11},{i12}} Meanwhile, Mn$_3$O$_4$ always exhibits insulating behavior with a band gap of about 2.0 eV.~\cite{{b1},{b2},{b3}}  

%---------------------------------------------------------------------
% Use the figure* environment if the figure should span across the
% entire page. There is no need to do explicit centering.
%---------------------------------------------------------------------

Mn$_3$O$_4$ is a challenging system to tackle by first-principles calculations, because strong electron correlations are combined with multiple degrees of freedom such as spin, orbital, charge, and lattice. Nevertheless, Mn$_3$O$_4$ has been extensively investigated through several computational methods in recent decades.~\cite{{c1},{c2},{c3},{c4}} Charter \textit{et al.} used Hartree-Fock calculations to estimate the band structure, magnetic ground state, and exchange coupling constants of Mn$_3$O$_4$.~\cite{c1} Although the Hartree-Fock method reproduced an insulating character, the electronic and magnetic structures were not consistent with experimental data; for example, the band gap was estimated to be about 10~eV. Franchini \textit{et al.} employed density functional theory (DFT) with Perdew-Burke-Ernzerhof (PBE), PBE \textit{plus} $U_{\mathrm{eff}}$ (PBE$+U_{\mathrm{eff}}$), and hybrid functionals to investigate the ground-state properties of Mn$_3$O$_4$.~\cite{c2} The ground state was found to be a half metal with ferrimagnetic ordering using the PBE$+U_{\mathrm{eff}}$ functional, whereas a ferrimagnetic insulator ground state was obtained within the hybrid functional. It was observed that PBE$+U_{\mathrm{eff}}$ favors a half-metallic state in reduced Mn oxides such as Mn$_3$O$_4$, Mn$_2$O$_3$, and MnO$_2$. Riberio \textit{et al.} further compared the ground states of Mn$_3$O$_4$ obtained using different hybrid functionals.~\cite{c3} Recently, Lim \textit{et al.} reported that DFT+$U$+$J$ reproduced a YK-FiM insulator with a band gap of 1 eV as a ground state.~\cite{c4} It was demonstrated that the DFT+$U$+$J$ calculation with an explicit exchange correction for the intraorbital Coulomb potential can accurately describe the ground state of reduced Mn oxides. Reliable first-principles calculations at DFT$+U$ level are of particular value because their low computational costs enable a variety of large-scale supercell calculations.

In this study, we systematically investigated the effects of anisotropic Coulomb interactions on the electronic and magnetic properties of Mn$_3$O$_4$ within the framework of DFT+$U$+$J$ calculations. We presents that the anisotropic interaction in Mn$^{3+}$ causes significant variations in various physical properties including the band gap, local magnetic moments, canted angle, spontaneous magnetic moment, and exchange coupling constants. Meanwhile, the anisotropic interaction in Mn$^{2+}$ is fully compensated without any noticeable effects. We found that the spin transfer into the Mn$^{3+}$ $d_{x^2-y^2}$ orbital is primarily tuned by the anisotropic interaction, triggering modification of the 90$^\circ$ correlation superexchange and noncollinear magnetic structure. These findings revealed why the anisotropic Coulomb correction is required in DFT$+U$ calculations for various reduced Mn oxides.

%---------------------------------------------------------------------
%\begin{widetext}
\section{Methodology and Computational Details}
\label{Methodology}
%\end{widetext}

DFT sometimes fails to describe the electronic structure of transition metal oxides with partially filled $d$ orbitals, due to the spurious self-interaction error that causes the localized $d$ state to be improperly destabilized.~\cite{{i13},{i14}} This error can be effectively corrected by employing Hubbard-$U$ type corrections for the localized orbitals, known as DFT$+U$. Two popular correction schemes have been being used for the DFT$+U$ approach. One is the spherically averaged scheme suggested by Dudarev~\cite{d1} (DFT$+U_{\mathrm{eff}}$) and the other is the rotationally invariant scheme proposed by Lichtenstein~\cite{d2} (DFT$+U$+$J$). In the DFT$+U_{\mathrm{eff}}$, the Coulomb interactions in the same $d$ orbital (direct Coulomb interactions) are corrected by a single $U_{\mathrm{eff}}$ parameter, and the interactions between the different $d$ orbitals (anisotropic Coulomb interactions) are set to be identical to an averaged value. In contrast, the direct and anisotropic Coulomb interactions are separately corrected by the Hubbard $U$ and exchange $J$ in the DFT+$U$+$J$. Nearly no significant differences have been observed for most materials regardless of which scheme was used. Recently, it was, however, shown that the anisotropic Coulomb interaction makes a critical difference in some multiorbital systems, including reduced Mn oxides, such as Mn$_3$O$_4$~\cite{c4}, Mn$_2$O$_3$~\cite{c4}, MnO$_2$~\cite{{c4},{e2},{e3}}, and LaMnO$_3$~\cite{e1}.

In the DFT+$U$+$J$ methodology, the DFT total energy is complemented not only by the direct and anisotropic Coulomb interactions, but also by considering the double-counted exchange energy correction, and thus given by~\cite{e1}
\begin{widetext}
\begin{equation}
\label{eq1}
  E_{\mathrm{DFT}+U+J} = E_\mathrm{DFT} + \frac{(U-J)}{2} {\sum_{i,\sigma}}^\prime f_{i\sigma}\left(1-f_{i\sigma}\right) 
   + \frac{1}{2} {\sum_{\substack{i,\sigma\\ j,\sigma^\prime}}}^\prime \left( C^{\sigma\sigma^\prime}_{ij} f_{i\sigma}f_{j\sigma^\prime}-\Delta X^{\sigma}_{ij}f_{i\sigma}f_{j\sigma^\prime}\delta_{\sigma\sigma^\prime}\right),
%\begin{split}
% & E_{\mathrm{DFT}+U+J} = E_\mathrm{DFT} + \frac{(U-J)}{2} \sum^\prime_{i,\sigma} \left( f_{i\sigma}-f^{2}_{i\sigma} \right) \\
%& + \frac{1}{2} \sum^\prime_{\substack{i,\sigma\\j,sigma^\prime}} \left( C^{\sigma\sigma^\prime}_{ij} f_{i\sigma}f_{j\sigma^\prime}-\Delta X^{\sigma}_{ij}f_{i\sigma}f_{j\sigma^\prime}\delta_{\sigma\sigma^\prime}\right),
%\end{split}
\end{equation}
\end{widetext}
where $\sum^\prime$ indicates the summation over only the Hubbard-$U$ corrected atoms, and $f_{i\sigma}$ is the eigenoccupation of the on-site density matrix with orbital and spin indices $i$ and $\sigma$. In this equation, the first term $E_\mathrm{DFT}$ is the total energy computed with electron-density based exchange-correlation functionals (PBE in this study), and the second term is the isotropic Coulomb correction part, the coefficient $(U-J)$ of which is identical to the effective Hubbard parameter $U_{\mathrm{eff}}$ used in the DFT$+U_{\mathrm{eff}}$ or Dudarev scheme.~\cite{d1} The third term is the anisotropic Coulomb correction part, in which two coefficients $C^{\sigma\sigma^\prime}_{ij}$ and $\Delta X^{\sigma}_{ij}$ are the specialized Coulomb and double-counting corrected exchange matrix elements for anisotropic interorbital interactions, respectively, defined in Ref.~[\onlinecite{e1}]. Note that, for isotropic cases, these matrices average to zero, being reduced to the DFT$+U_{\mathrm{eff}}$ scheme, but for anisotropic cases, on the other hand, these correction matrices are regarded as additional Coulombic and exchange corrections induced from the angular characters of interactions. Thus, the $U$+$J$ correction $\Delta\epsilon_{i\sigma}$ to the DFT energy eigenvalue $\epsilon_{i\sigma}^\mathrm{DFT}$ can be evaluated by taking the derivative of the correction parts in Eq.~(\ref{eq1}) with respect to $f_{i\sigma}$ as 
\begin{align}
  \Delta\epsilon_{i\sigma} &=\frac{\partial{\left(E_{\mathrm{DFT}+U+J}-E_\mathrm{DFT}\right)}}{\partial f_{i\sigma}} \nonumber\\
  &=(U-J)\left(\frac{1}{2}-f_{i\sigma}\right)+\sum_{j\sigma^\prime}\left(C_{ij}^{\sigma\sigma^\prime}f_{j\sigma^\prime}-\Delta X_{ij}^\sigma f_{j\sigma^\prime}\delta_{\sigma\sigma^\prime}\right).
  \label{eq2}
\end{align}
Eq.~(\ref{eq2}) can be expressed in a more compact vector-matrix multiplications, especially for anisotropic $d$ orbital systems, whose correction matrices $C^{\sigma\sigma^\prime}$ and $\Delta X^{\sigma}$ are proportional to the exchange parameter $J$, as~\cite{e1}
\begin{equation}
\label{eq3}
  \Delta\epsilon_\sigma=(U-J)\left(\frac{1}{2}-f_\sigma\right)+J\left(A^\sigma f_\sigma+B^\sigma f_{\bar{\sigma}}\right),
%\begin{split}
%E_{\mathrm{DFT}+U+J} = & E_\mathrm{DFT} + \frac{(U-J)}{2} \sum_{at,\sigma,i} \left( f_{i\sigma}-f^{2}_{i\sigma} \right) \\
%& + JA^{\sigma}f_{i\sigma}f_{j\sigma '}-JB^{\sigma}f_{i\sigma}f_{j\sigma}\delta_{\sigma\sigma '}
%\end{split}.
\end{equation}
where $\bar{\sigma}$ denotes the opposite spin to $\sigma$, and $A^\sigma=\left(C^{\sigma\sigma}-\Delta X^\sigma\right)/J$ and $B^\sigma=C^{\sigma\bar{\sigma}}/J$ are dimensionless matrices. In terms of a basis composed of the $e_g$ ($d_{x^2-y^2}$ and $d_{3z^2-r^2}$) and $t_{2g}$ ($d_{xy}$, $d_{yz}$, and $d_{zx}$) orbitals, the matrix elements of these dimensionless matrices can be evaluated to be~\cite{e1}
\begin{equation*}
A^{\sigma} =\begin{pmatrix*}[r]
   0.00 & -0.52 &  0.86 & -0.17 & -0.17 \\ 
  -0.52 &  0.00 & -0.52 &  0.52 &  0.52 \\
   0.86 & -0.52 &  0.00 & -0.17 & -0.17 \\
  -0.17 &  0.52 & -0.17 &  0.00 & -0.17 \\
  -0.17 &  0.52 & -0.17 & -0.17 &  0.00 
\end{pmatrix*}
\end{equation*}
and
\begin{equation*}
B^{\sigma} =\begin{pmatrix*}[r]
   1.14 & -0.63 &  0.29 & -0.40 & -0.40 \\
  -0.63 &  1.14 & -0.63 &  0.06 &  0.06 \\
   0.29 & -0.63 &  1.14 & -0.40 & -0.40 \\
  -0.40 &  0.06 & -0.40 &  1.14 & -0.40 \\
  -0.40 &  0.06 & -0.40 & -0.40 &  1.14 
\end{pmatrix*}.\\
\end{equation*}

DFT+$U$+$J$ calculations were carried out using the Vienna \textit{ab initio} simulation package (VASP) code.~\cite{f1} We used the PBE functional~\cite{f2} for the exchange-correlation functional, and the projected-augmented-wave (PAW) method~\cite{f3} was applied to describe the potential of the core electrons. We defined the $U$ and $J$ parameters separately for Mn$^{2+}$ and Mn$^{3+}$ to investigate the role of anisotropic interactions at each site. The energy cutoff for the plane-wave basis set was 500~eV and a $\Gamma$-centered $6\times6\times4$ Monkhorst-pack grid was used for sampling the Brillouin zone. To focus on the effects of anisotropic Coulomb interactions, we fixed the lattice parameters to the experimental values for all calculations, i.e $a=b=5.76$~{\AA} and $c=9.46$~\AA.~\cite{a2} In this study, we only considered the experimental YK-FiM phase, which is the ground state of Mn$_3$O$_4$ verified also in the DFT+$U$+$J$ scheme.~\cite{c4}

\section{Results}
\label{Results}

Exchange parameter $J$ plays two different roles in the Hubbard correction of the band energy eigenvalues as revealed in Eq.~(\ref{eq3}). On the one hand, the strength of the direct Coulomb repulsion is reduced by $J$ as in $(U-J)$, which is the same as $U_\mathrm{eff}$ in the Dudarev scheme.~\cite{d1} On the other hand, $J$ contributes an additional orbital splitting as a direct consequence of the interorbital interactions. By comparing the numerical results computed with different $J$ while keeping $(U-J)$ constant, we can solely focus on the role of anisotropic interactions in the electronic and magnetic properties of Mn$_3$O$_4$. In other words, we scanned the Hubbard-$U$ parameter space with the axes $(U-J)^{2+}$, $(U-J)^{3+}$, $J^{2+}$, and $J^{3+}$, not the axes $U^{2+}$, $U^{3+}$, $J^{2+}$ and $J^{3+}$. 

\subsection{Electronic structures}
\label{Electronic structures}

%---------------------------------------------------------------------
% Use the figure* environment if the figure should span across the
% entire page. There is no need to do explicit centering.
\begin{figure}[t]
\includegraphics[width=1\linewidth]{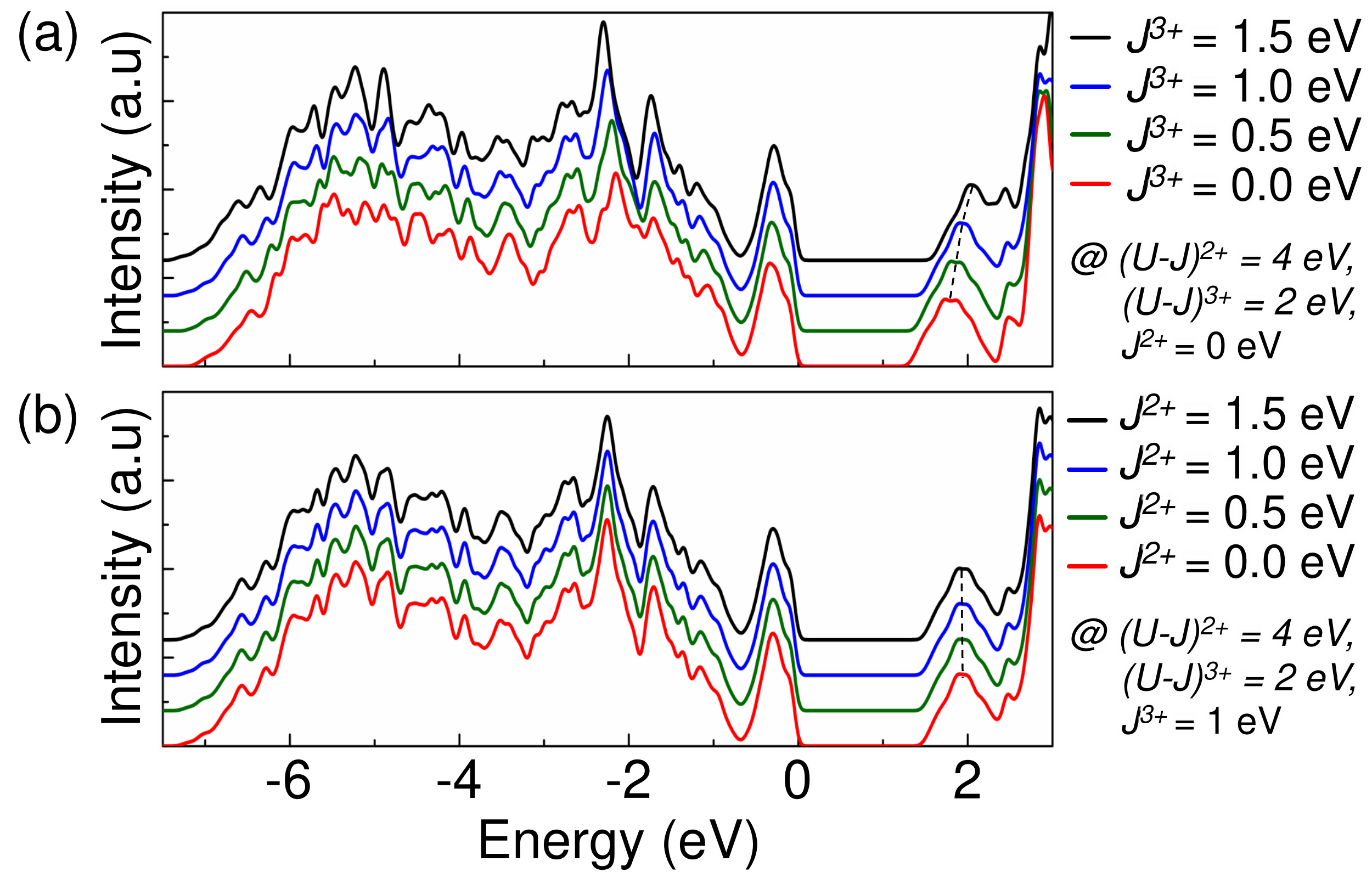}
\caption{Densities of states (DOSs) of YK-FiM Mn$_3$O$_4$ calculated for fixed values of $(U, -J)^{2+}=4$~eV and $(U-J)^{3+}=2$~eV (a) with varying $J^{3+}$ values and a fixed value of $J^{2+}=0$~eV, and (b) with varying $J^{2+}$ and a fixed value of $J^{3+}=1$~eV. \label{Fig2}}
\end{figure}
%---------------------------------------------------------------------

For typical parameter values of $(U-J)^{2+}=4$~eV and $(U-J)^{3+}=2$~eV, we first calculated densities of states (DOSs) of Mn$_3$O$_4$ with different values of the $J^{2+}$ and $J^{3+}$ parameters. It turned out that the anisotropic Coulomb interaction in Mn$^{3+}$ plays a more significant role in orbital energies than that in Mn$^{2+}$, increasing the band gap from 1.3~eV with $J^{3+}=0$~eV to 1.6~eV with $J^{3+}=1.5$~eV, as shown in Fig.~\ref{Fig2} (a). This suggests that the anisotropic interaction in Mn$^{3+}$ have more considerable effects on superexchange coupling in Mn$_3$O$_4$. In contrast, the anisotropic interactions in Mn$^{2+}$ did not cause any change in the orbital energies. The insensitivity of Mn$^{2+}$ is attributed to their orbital occupations. The Mn$^{2+}$ in Mn$_3$O$_4$ have a high-spin $d_5$ orbital occupation with majority-spin orbitals fully occupied while its minority-spin orbitals are totally empty. In this electron configuration, the anisotropic Coulomb interactions are fully compensated, and the orbital energies are not affected at all by the anisotropic interactions. This result further verifies why the anisotropic Coulomb interactions are particularly important for the reduced Mn oxides, but not for MnO.~\cite{{c2},{h1},{h2},{h3}}

\subsection{Magnetic properties}
\label{Magnetic properties}

%---------------------------------------------------------------------
% Use the figure* environment if the figure should span across the
% entire page. There is no need to do explicit centering.
\begin{figure*}[t]
\includegraphics[width=1\linewidth]{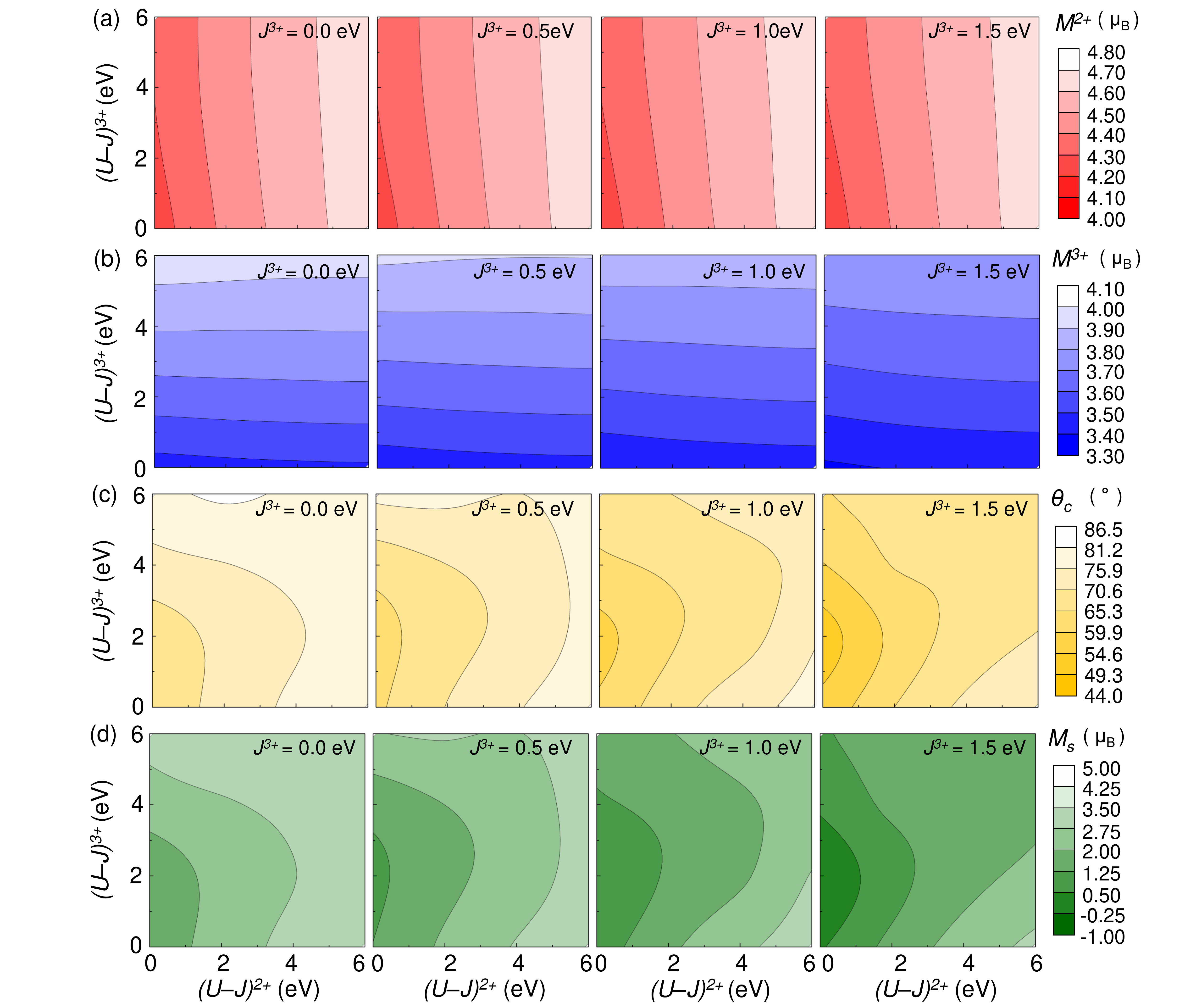}
\caption{Contour maps for local magnetic moments (a) $M^{2+}$ and (b) $M^{3+}$, respectively, of Mn$^{2+}$ and Mn$^{3+}$, (c) canted angle $\theta_{c}$ defined in Fig.~\ref{Fig1} (b), and (d) spontaneous magnetic moments $M_{s}$ of YK-FiM Mn$_3$O$_4$ evaluated with different sets of Hubbard parameters. Each map displays its corresponding magnetic property as a function of the effective Hubbard parameters $(U-J)^{2+}$ and $(U-J)^{3+}$ at a specific $J^{3+}$ with $J^{2+}=0$~eV. The contour maps for each quantity are color-coded according to the values given in the scale bar.
\label{Fig3}}
\end{figure*}
%---------------------------------------------------------------------

Several magnetic observables of the YK-FiM Mn$_3$O$_4$ were evaluated while adjusting Hubbard parameters. These observables were represented in a color-coded contour map with $(U-J)^{2+}$ and $(U-J)^{3+}$ as the $x$- and $y$-axes while changing $J^{3+}$ values from 0.0~eV to 1.5~eV
For a given specific $J^{3+}$ value, each was displayed in color-coded contour maps as a function different sets
Figure~\ref{Fig3} summarizes the magnetic observables of YK-FiM Mn$_3$O$_4$ calculated with different sets of Hubbard parameters. , but keeping $J^{2+}$ fixed to a specific value (here, 0~eV) due to its insensitivity verified in the band structure shown in Fig.~\ref{Fig2} (b). Figures~\ref{Fig3} (a) and (b) present the local magnetic moments $M^{2+}$ and $M^{3+}$, respectively, of Mn$^{2+}$ and Mn$^{3+}$. It is shown that the former increases as $(U-J)^{2+}$ increases, while the latter does as $(U-J)^{3+}$ is elevated. This is consistent with the previous knowledge that the direct Coulomb repulsion enhances the spin- and orbital-polarizations. Moreover, while $M^{2+}$ appears relatively independent of the exchange $J^{3+}$ values, $M^{3+}$ tends to decrease as the exchange $J^{3+}$ increases, as displayed in Fig.~\ref{Fig3} (a) and (b). This indicates that the anisotropic interaction in Mn$^{3+}$ reduces its local magnetic moments. Additionally, the canted angle and the spontaneous magnetic moment are also modified by the anisotropic interaction in Mn$^{3+}$, as shown in Figs.~\ref{Fig3}(c) and (d). Both the canted angle and the spontaneous magnetic moment decrease monotonically as the exchange $J^{3+}$ increases. Our results reveal that the anisotropic Coulomb interactions in Mn$^{3+}$ are the crucial factor in determining the magnetic ground state of Mn$_3$O$_4$.

%---------------------------------------------------------------------
% Use the figure* environment if the figure should span across the
% entire page. There is no need to do explicit centering.
\begin{figure}[t]
\includegraphics[width=1\linewidth]{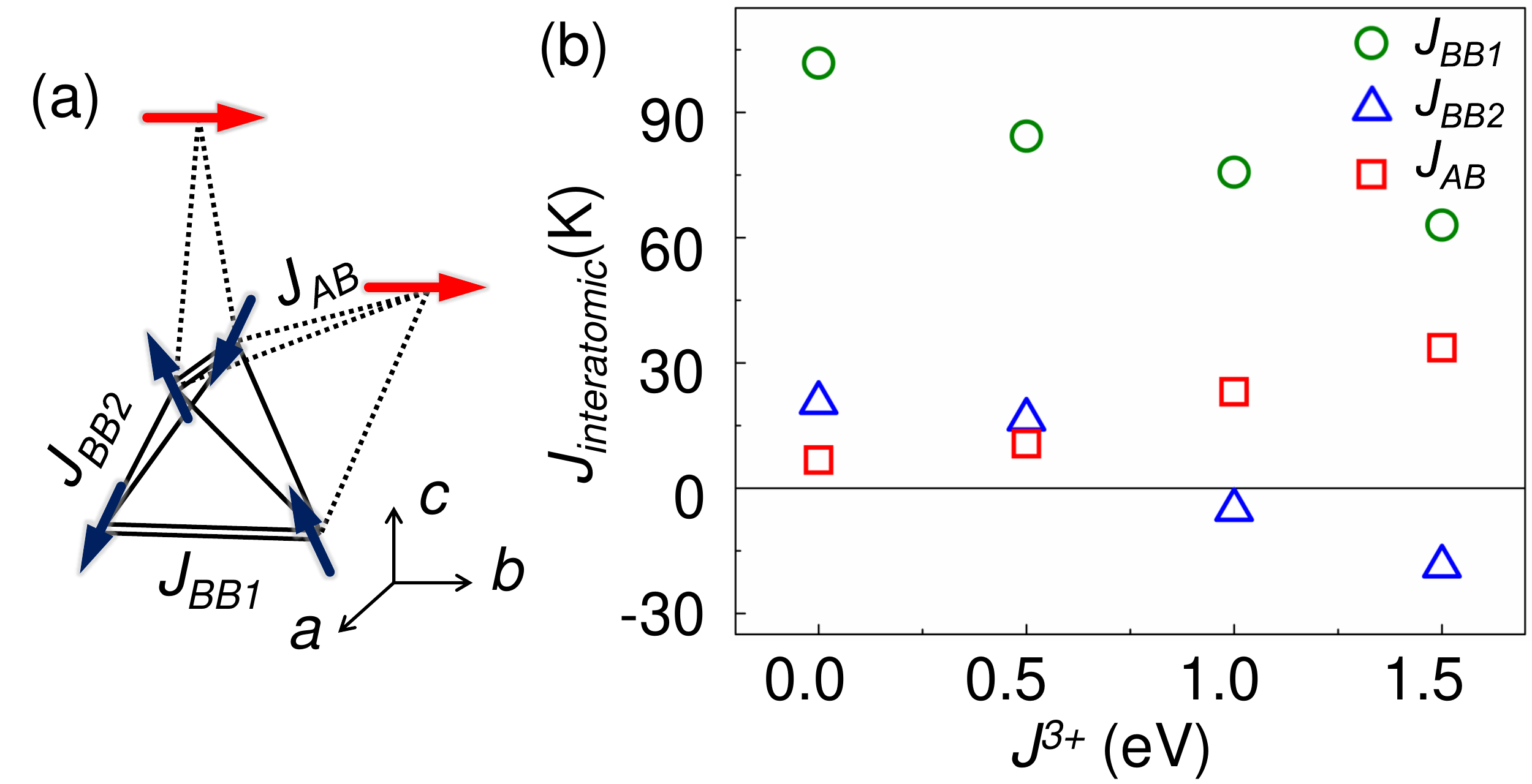}
\caption{(a) Different interatomic exchange couplings $J_\mathrm{BB1}$, $J_\mathrm{BB2}$, and $J_\mathrm{AB}$ marked in the schematic spin configuration of the YK-FiM Mn$_3$O$_4$ shown in Fig.~\ref{Fig1} (b). $J_\mathrm{BB1}$ indicates the short-distance interaction between two neighboring octahedral B-site ions within the same chain along either the $a$ or $b$ direction, whereas $J_\mathrm{BB2}$ represents the long-distance interchain interaction between two perpendicular octahedral chains. $J_\mathrm{AB}$ denotes the exchange coupling constant between neighboring tetrahedral A- and octahedral B-site ions. (b) Three exchange coupling constants $J_\mathrm{BB1}$, $J_\mathrm{BB2}$, and $J_\mathrm{AB}$ as a function of the exchange parameter $J^{3+}$ verifying that their dependence on the anisotropic Coulomb interactions of Mn$^{3+}$.
\label{Fig4}}
\end{figure}
%---------------------------------------------------------------------

%Remarkably, the exchange coupling constants in Mn$_3$O$_4$ are strongly affected by the anisotropic interactions in Mn$^{3+}$. 
We further explored the effect of $J^{3+}$ on three different interatomic exchange couplings of Mn$_3$O$_4$ classified as follows. As depicted in Fig.~\ref{Fig4}(a), $J_\mathrm{BB1}$ indicates the short-distance interaction between two neighboring octahedral B-site ions located within the same chain along either the $a$ or $b$ direction, whereas $J_\mathrm{BB2}$ represents the long-distance interchain interaction between two perpendicular octahedral chains. $J_\mathrm{AB}$ denotes the exchange coupling constant between a neighboring tetrahedral A- and an octahedral B-site ions. This set of Hubbard parameters was determined as the optimum values for describing the magnetic properties of Mn$_3$O$_4$, which will be discussed in the following paragraph. The exchange coupling constants were extracted by mapping the energies of various magnetic disorders onto a classical Heisenberg Hamiltonian
%\begin{widetext}
\begin{equation*}
  H=J_\mathrm{BB1}\sum_\mathrm{BB1}\mathbf{S}_i{\cdot}\mathbf{S}_j
   +J_\mathrm{BB2}\sum_\mathrm{BB2}\mathbf{S}_i{\cdot}\mathbf{S}_j
   +J_\mathrm{AB}\sum_\mathrm{AB}\mathbf{S}_i{\cdot}\mathbf{S}_j,
\end{equation*}
%\end{widetext}
where $\mathbf{S}_i{\cdot}\mathbf{S}_j$ is the inner product between two spins localized at $i$ and $j$ sites.~\cite{h2} Each summation is allowed only over the pairs of spins constrained by its corresponding subscript. These three fitted exchange coupling constants are presented as a function of $J^{3+}$ with constant values of $(U-J)^{2+}=4$~eV, $(U-J)^{3+}=2$~eV, and $J^{2+}=0$~eV in  
Fig.~\ref{Fig4}(b). It was shown that $J_\mathrm{BB1}$, which is always positive indicating the antiferromagnetic (AFM) intrachain interaction, is notably attenuated by the anisotropic interaction, while the interchain exchange coupling $J_\mathrm{BB2}$ induces the transformation from AFM to ferromagnetic (FM) states with the anisotropic interaction $J^{3+}$ as indicated by its sign change. On the other hand, $J_\mathrm{AB}$ becomes stronger as the anisotropic interaction strengthened, maintaining the AFM interaction between A- and B-site ions. The strong AFM $J_\mathrm{BB1}$ and the weak FM $J_\mathrm{BB2}$ are the key features of magnetic interactions in Mn$_3$O$_4$, measured by inelastic neutron scattering experiments.~\cite{i11} Our results demonstrate that this key characteristic is determined by the anisotropic Coulomb interactions in Mn$^{3+}$.

\subsection{Optimal parameters of Hubbard $U$ and exchange $J$ for Mn$_3$O$_4$}
\label{Parameters}

%---------------------------------------------------------------------
\begin{table*}[b]
\label{table1}
\caption{Various magnetic properties of the Yafet-Kittel phase of Mn$_{3}$O$_{4}$ evaluated with $(U^{2+},J^{2+})=(4,0)$~eV and $(U^{3+},J^{3+})=(3,1)$~eV. The corresponding experimental values are given for comparison.~\cite{{a3},{i11}}}
\label{Table1}
\begin{ruledtabular}
\begin{tabular}{cccccccc}
   & $M^{2+}$ ($\mu_{B}$) & $M^{3+}$ ($\mu_{B}$) & $M^\mathrm{spon}$ ($\mu_{B}$) & $\theta^\mathrm{cant}$ ($^{\circ}$) & $J_\mathrm{BB1}$ (K) & $J_\mathrm{BB2}$ (K) & $J_\mathrm{AB}$ (K)\\
 \hline
 \mbox{This work} & 4.56 & 3.61 & 1.77 & 67.6 & 75.7 & -5.2 & 23.08 \\
 \mbox{Experiments} & 4.57 & 3.51 & 1.84 & 66.9 & 55.1 & -3.2 & 6.3 \\
\end{tabular}
\end{ruledtabular}
\end{table*}
%---------------------------------------------------------------------

With $(U-J)^{2+}=4$~eV, $(U-J)^{3+}=2$~eV, $J^{2+}=0$~eV, and $J^{3+}=1$~eV, we applied the DFT+$U$+$J$ scheme to evaluate various magnetic properties, such as local magnetic moments ($M^{2+}$ and $M^{3+}$), spontaneous magnetic moment ($M^\mathrm{spon}$), canted angle ($\theta^\mathrm{cant}$), and exchange coupling constants ($J_\mathrm{BB1}$, $J_\mathrm{BB2}$, and $J_\mathrm{AB}$). As summarized in Table~\ref{Table1} together with the corresponding experimental data available,~\cite{{a3},{i11}} our DFT+$U$+$J$ calculations accurately reproduced its magnetic moments and canted angle observed experimentally. Importantly, the weak FM interchain superexchange was also predicted by this set of Hubbard parameters. The optimal Hubbard parameters were different for Mn$^{2+}$ and Mn$^{3+}$, suggesting that the effective Coulomb interaction that acts on $d$ electrons may vary depending on the oxidation states, local symmetry, and/or hybridization. Specifically, the direct Coulomb interaction in Mn$^{2+}$ ($(U-J)^{2+}=4$~eV) is stronger than that in Mn$^{3+}$ ($(U-J)^{3+}=2$~eV). The reduced Coulomb interaction in Mn$^{3+}$ can be attributed to its strong charge-transfer or hybridization character.~\cite{g1} In this study, we considered separate Hubbard parameters for Mn$^{2+}$ and Mn$^{3+}$ to clarify the role of each on-site interaction in the electronic and magnetic properties of Mn$_3$O$_4$. It may be difficult to define the Hubbard parameters separately when investigating the physical properties of non-bulk configurations (such as surfaces, interfaces, and defects). In this case, $U=4.0$~eV and $J=1.2$~eV, as employed by Lim \textit{et al.}, would be a reasonable choice for a single set of Hubbard parameters for Mn$_3$O$_4$.~\cite{c4} However, the weak FM interchain interactions are not reproduced in calculations using a single set of Hubbard parameter, although the overall results are qualitatively consistent with our results.~\cite{c4}

\section{Discussions}
\label{Discussions}

%---------------------------------------------------------------------
% Use the figure* environment if the figure should span across the
% entire page. There is no need to do explicit centering.
\begin{figure}[t]
\includegraphics[width=1\linewidth]{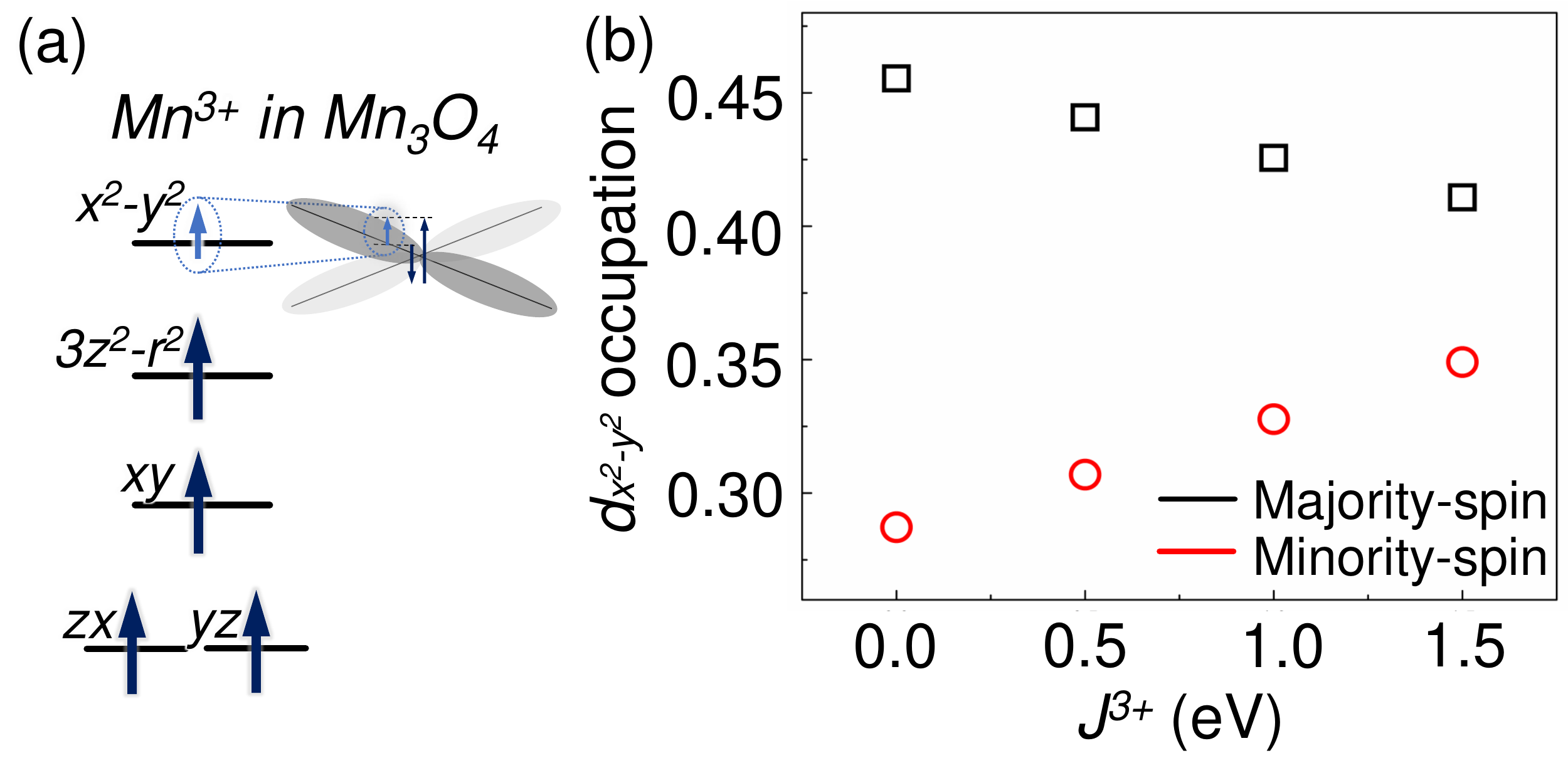}
\label{fig5}
\caption{(a) Schematic of the $d$ orbital levels and the orbital occupations of an Mn$^{3+}$ ion in Mn$_3$O$_4$. The inset explains the partially occupied $d_{x^2-y^2}$ orbital, which attributed to the spin transfer from O$^{2-}$ through the correlation superexchange. (b) $d_{x^2-y^2}$ orbital occupations of the majority- and minority spins of an Mn$^{3+}$ ion as a function of exchange parameter $J^{3+}$.
\label{Fig5}}
\end{figure}
%---------------------------------------------------------------------

The eigenoccupations of the on-site $d$ orbital density matrix ($f_{\sigma}$) are fictitious auxiliary physical quantities, but they can often provide a direct insight into the orbital states of the transition metal ions.~\cite{{e1},{e3}} With $(U-J)^{2+}=4$~eV, $(U-J)^{3+}=2$~eV, $J^{2+}=0$~eV, and $J^{3+}=0$~eV, the eigenoccupations of the A- and B-site Mn atoms were calculated to be
%
%\begin{widetext}
\begin{equation*}
 \left ( f_{\sigma} | f_{\bar{\sigma}} \right ) = \begin{pmatrix}
0.940 & 0.018 \\ % d_{x^2-y^2}
0.943 & 0.025 \\ % d_{3z^2-r^2}
0.944 & 0.062 \\ % d_{xy}
0.948 & 0.059 \\ % d_{yz}
0.944 & 0.061 \\ % d_{zx}
\end{pmatrix}
\;\mbox{and}\;
\left ( f_{\sigma} | f_{\bar{\sigma}} \right ) = \begin{pmatrix}
0.456 & 0.287 \\ % d_{x^2-y^2}
0.960 & 0.142 \\ % d_{3z^2-r^2}
0.943 & 0.067 \\ % d_{xy}
0.951 & 0.082 \\ % d_{yz}
0.939 & 0.057 \\ % d_{zx}
\end{pmatrix},
\end{equation*}
%\end{widetext}
%
respectively, where the occupations of the $d_{x^2-y^2}$, $d_{3z^2-r^2}$, $d_{xy}$, $d_{yz}$, and $d_{zx}$ orbitals are represented from top to bottom. The A-site Mn$^{2+}$ ion was confirmed to have a high-spin (HS) $d_5$ orbital configuration (HS Mn$^{2+}$) with all majority-spin orbitals almost fully occupied but minority-spin orbitals nearly empty. For the B-site Mn$^{3+}$ ion, on the other hand, the four lower energy $d$ orbitals are fully occupied for majority-spin $d_{3z^2-r^2}$, $d_{xy}$, $d_{yz}$, and $d_{zx}$ orbitals, which is consistent with the HS Mn$^{3+}$ configuration. Interestingly, the majority- and minority-spin $d_{x^2-y^2}$ orbitals, which are expected to be fully empty in the ideal Mn$^{3+}$ configuration, were partially occupied in Mn$_3$O$_4$, which is attributed to the charge transfer from O$^{2-}$ through the correlation superexchange, as depicted in Fig.~\ref{Fig5}(a). Furthermore, the amount of occupation of the majority-spin $d_{x^2-y^2}$ orbitals is different from that of the minority-spin $d_{x^2-y^2}$ orbitals, resulting in a net spin moment in $d_{x^2-y^2}$ orbital. 

We explored how the orbital occupation of Mn$^{3+}$ varies with the strength of its anisotropic interaction. The occupations of the four lower occupied orbitals rarely depend on the anisotropic Coulomb interaction, whereas that of the $d_{x^2-y^2}$ orbital is systematically modified, as shown in Fig.~\ref{Fig5}(b). Its majority- and minority-spin orbitals become less and more occupied with the anisotropic interactions, respectively, resulting in the reduction of the net spin moment, which can be interpreted as the spin transfer into the $d_{x^2-y^2}$ orbital, but its total occupation remains. The reduced net spin moment is in-line with the reduction in the local magnetic moments $M^{3+}$ of Mn$^{3+}$ displayed in Fig.~\ref{Fig3}(b).

The magnetic interaction between two neighboring Mn$^{3+}$ ions, which are connected via an O$^{2+}$ ion to each other at $90^\circ$, is recognized to be governed by two different exchange couplings.~\cite{{g2},{i11}} One is the direct overlap between $d_{xy}$ orbitals, causing the short-distance intrachain interaction $J_{BB1}$ to be very strongly AFM. The other is the $90^\circ$ correlation superexchange referring to the magnetic interaction whereby two electrons of an O$^{2-}$ ion relay the exchange coupling between neighboring cations.~\cite{g2} Importantly, the strongest spin channel for the correlation superexchange always involves the $d_{x^2-y^2}$ orbitals, and the anisotropic Coulomb interaction determines that spin channel. The anisotropic interaction reduces the AFM interaction in $J_{BB1}$ due to the weakened correlation superexchange by a shrunk spin channel. Indeed, the sign of the 90$^\circ$ correlation superexchange is very sensitive to the characteristics of the broker electrons in the O$^{2-}$ ions.~\cite{g2} If two electrons relaying the 90$^\circ$ correlation superexchange are from the same orbital, the magnetic interaction remains AFM keeping the sign, but if they are from different (and orthogonal) orbitals, the interaction may become FM through the sign change. That is, the orbital and hybridization state of O$^{2-}$ determine the sign of the 90$^\circ$ correlation superexchange. The reduced net spin transfer may have perturbed the hybridized state of the O$^{2-}$ ions, yielding the transition of $J_{BB2}$ from weak AFM to weak FM.

\section{Conclusions}
\label{Conclusions}

We analyzed the role of the anisotropic Coulomb interaction playing in the electronic and magnetic ground state of mixed-valent Mn$_3$O$_4$ within the framework of DFT+$U$+$J$ with separate sets of distinct Hubbard parameters for Mn$^{2+}$ and Mn$^{3+}$. The anisotropic interaction $J^{3+}$ of the Mn$^{3+}$ ion extensively affects a variety of ground-state properties such as the band gap, local magnetic moments, canted angle, spontaneous magnetic moments, and superexchange coupling, while $J^{2+}$ the counterpart of the Mn$^{2+}$ ion is fully compensated without any noticeable changes. It is importantly found that the 90$^\circ$ correlation superexchange is characterized by the spin transfer into the $d_{x^2-y^2}$ orbital of the Mn$^{3+}$ ion, which can be controlled by the anisotropic interaction in Mn$^{3+}$. The DFT+$U$+$J$ calculations with the sizable anisotropic interaction $J^{3+}$ reproduce the weak FM interchain interactions, which has not been obtained in previous calculations. The best description was achieved when $U^{2+}=4$~eV, $J^{2+}=0$~eV, $U^{3+}=3$~eV, and $J^{3+}=1$~eV. Our study provides an important insight into why the exchagne parameter $J$ should be explicitly taken in account in Hubbard corrections, especially for reduced Mn oxides such as Mn$_3$O$_4$, Mn$_2$O$_3$, MnO$_2$, and LaMnO$_3$. The anisotropic interactions strongly modify the spin transfer into unoccupied $d$ orbitals, leading to significant corrections of intraatomic superexchange interactions.  

\section*{Acknowledgements}
We gratefully acknowledge financial support from the South Korean government (MSIT) through the National Research Foundation (NRF) of Korea (No. 2019R1A2C1005417). Some portion of our computational work was done using the resources of the KISTI Supercomputing Center (KSC-2020-CRE-0011).

%+++++++++++++++++++++++++++++++++++++++++++++++++++++++++++++++++++++
\bibliographystyle{apsrev}% your bst file here
\bibliography{main.bib} %your bib file here
\end{document}